\documentclass[12pt]{article}

\usepackage{amsfonts,amsthm,dsfont,stmaryrd,amsmath,amsopn,amssymb}

\newcommand{\defn}[1]{\emph{#1}}

\newcommand{\N}{\mathds{N}}
\newcommand{\C}{\mathds{C}}

\newcommand{\bcup}{\bigcup}
\newcommand{\ls}{\leq^*}

\newcommand{\stem}{\lhd}
\newcommand{\bs}{\bigskip\noindent}
\DeclareMathOperator{\id}{id}

\theoremstyle{plain}
\newtheorem{theorem}{Theorem}[section]
\newtheorem{lemma}[theorem]{Lemma}
\newtheorem{corollary}[theorem]{Corollary}
\newtheorem{proposition}[theorem]{Proposition}

\theoremstyle{definition}

\newtheorem{construction}[theorem]{Construction}

\theoremstyle{remark}
\newtheorem{remark}[theorem]{Remark}

\usepackage{enumerate,fullpage}

\newcommand{\A}{\mathcal{A}}
\newcommand{\B}{\mathcal{B}}
\newcommand{\R}{\mathcal{R}}
\newcommand{\Q}{\mathcal{Q}}
\renewcommand{\C}{\mathcal{C}}
\newcommand{\CL}{\mathfrak{C}}
\newcommand{\CLfin}{\mathfrak{C}_{\hbox{fin}}}
\newcommand{\CLrel}{\mathfrak{C}_{\hbox{rel}}}
\newcommand{\U}{\mathcal{U}}
\renewcommand{\S}{\mathcal{S}}
\DeclareMathOperator{\past}{past}
\DeclareMathOperator{\Aut}{Aut}

\begin{document}

\title{Universal homogeneous causal sets}
\author{Manfred Droste\\Institut f\"ur Informatik\\Universit\"at Leipzig\\D-04109 Leipzig, Germany\\email: droste@informatik.uni-leipzig.de}
\maketitle
\begin{abstract}
Causal sets are particular partially ordered sets which have been
proposed as a basic model for discrete space-time in quantum gravity.
We show that the class $\CL$ of all countable past-finite causal sets
contains a unique causal set $(U,\leq)$ which is universal (i.e., any member of $\CL$
can be embedded into $(U,\leq)$) and homogeneous (i.e., $(U,\leq)$ has
maximal degree of symmetry). Moreover, $(U,\leq)$ can be constructed both
probabilistically and explicitly.
\end{abstract}

\section{Introduction}

The causal set hypothesis asserts that the ultimate structure of
space-time in quantum gravity is discrete, and that a fundamental
relationship between points in space-time is causality, enabling us
to say that $x$ is in the past of $y$. Such a structure can be
naturally modelled by partially ordered sets $(S,\leq)$ which are
locally finite (i.e., any interval $[x,y]$ between two points $x,y$
is finite) or which are even past-finite (i.e., the past of each point
$x$ is finite). Locally finite partially ordered sets are called
causal sets and have been investigated in detail, cf. e.g.
\cite{S02}. Moreover, an interesting sequential growth
dynamics for finite causal sets was investigated in \cite{RS00,BDGHS01,BDGHS03}:
a finite causal set can be extended by a single
element $z$ by adding it to the given partial order as a new maximal
element (naturally, $z$ cannot be in the past of any given element).
Continuing in this way, one obtains a countable past-finite causal set.
Conversely, any countable past-finite causal set can be built up this
way.
 For further work on this dynamics, see e.g. \cite{AM03,G03,MORS01}

It is the goal of this paper to investigate the class of countable
past-finite causal sets. We will show that there is a countable
past-finite causal set $(U,\leq)$ which is universal and homogeneous.
Here, \emph{universal} means that $(U,\leq)$ contains an isomorphic
copy of any countable past-finite causal set as a natural substructure.
\emph{Homogeneous} means that any isomorphism between two finite
substructures (stems) of $(U,\leq)$ extends to an automorphism of
$(U,\leq)$. Thus homogeneity intuitively says that $(U,\leq)$ has
the highest possible degree of structural symmetry. Moreover, with
these two properties, universality and homogeneity, $(U,\leq)$ is
unique up to isomorphism in the class of all countable past-finite
causal sets. Our proof employs the Fra\"{\i}ss\'{e}-J\'{o}nsson
theorem well-known in model theory for constructions of homogeneous
relational structures. We also 
give an explicit order-theoretic construction of the universal homogeneous past-finite causal set, and we 
show that the larger class of all
countable causal sets (not requiring 'past-finite'), in contrast, does
not contain a universal object.

In our second result, we will describe a probabilistic construction
of the universal homogeneous past-finite causal set $(U,\leq)$.
Probabilistic procedures for constructing one-point extensions of
finite causal sets were crucial for \cite{RS00,G03} and
investigated in detail. Here we also propose a probabilistic
construction. It is motivated by a classical construction of Erd\"os
and R\'{e}nyi \cite{ER63} of the random graph (which is universal
and homogeneous in the class of all countable graphs). More
recently, a similar construction of the universal homogeneous
countable partial order was given in \cite{DK03}. If we employ the
present probabilistic one-point extensions of finite causal sets
successively, we obtain an infinite sequence of causal sets whose
union is a countable past-finite causal set, the \defn{random} past-finite
causal set. Then we show that with probability 1, this random causal
set is universal and homogeneous (hence unique up to isomorphism).
Our construction of one-point extensions differs from the ones
investigated in \cite{RS00} in technical details, which we discuss briefly, and 
it would be interesting to investigate the relationship further.
%
Finally, we describe a simple explicit number-theoretic construction
of $(U,\leq)$. This is motivated by a classical construction of the
random graph \cite{Rad64}. 
We close with a short discussion.
\section{Universal causal sets}

First we establish some terminology. For a fuller introduction to
causal sets, see \cite{S02}.

A partially ordered set  (poset) is a pair $(S,\leq)$ where $S$ is a
set and $\leq$ is a binary relation on $S$ which is reflexive,
antisymmetric and transitive. Let $(S,\leq)$ be a poset. For $x,y\in
S$ we write $x<y$ to denote that $x\leq y$ and $x\neq y$. If  
 $x<y$, we let $[x,y]=\{s\in S\mid x\leq s\leq y\}$, and for
$x\in S$ let $\past_S(x)=\{s\in S\mid s\leq x\}$. Then $(S,\leq)$ is
called \defn{locally finite}, if each interval $[x,y]$ $(x,y\in
S,x<y)$ is finite, and \defn{past-finite}, if $\past_S(x)$ is finite for
each $x\in S$. Clearly, any past-finite poset is locally finite (but
not conversely as seen by the set of negative integers with their
natural order). A \defn{causal set} (or \defn{causet}) is a locally
finite poset. If $X\subseteq S$ is a subset, let
$\past_S(X)=\bcup_{x\in X}\past(x)$. If $S$ is clear from the context,
we also write $\past(x)$ and $\past(X)$. Then $X$ is called a \defn{stem
of $S$}, denoted $(X,\leq)\stem(S,\leq)$ or $X\stem S$, if
$\past(X)\subseteq X$ (note that here, in slight generalization of
\cite{S02}, we do not require $X$ to be finite).

In \cite{RS00,S02}, constructions of countable causal sets $(S,\leq)$ are
give as unions of chains of finite posets
$(A_1,\leq_1)\stem(A_2,\leq_2)\stem\dots$. Indeed, given such a
chain, let $S=\bcup_{i\in\N}A_i$, and for $x,y\in S$ put $x\leq y$
if for some $i\in\N$ we have $x,y\in A_i$ and $x\leq_i y$. Then
$(S,\leq)$ is a past-finite countable poset and $A_i\stem S$ for each
$i\in\N$. Such constructions will also be crucial for this paper.

Let $(S,\leq)$ and $(S',\leq)$ be two posets. A mapping $f:S\to S'$
is called an (\defn{order}-) \defn{embedding of $(S,\leq)$ into
$(S',\leq)$}, if for any $x,y\in S$ we have $x\leq y$ iff $f(x)\leq
f(y)$. Clearly, any embedding is one-to-one. An embedding
$f:(S,\leq)\to(S',\leq)$ is called an \defn{isomorphism}, if $f$ is
onto, and a \defn{stem-embedding}, if $f(S)\stem S'$. So, $f$ is a
stem-embedding iff $f$ is an isomorphism from $(S,\leq)$ onto a stem
of $(S',\leq)$. An isomorphism of $(S,\leq)$ onto itself is called
an \defn{automorphism} of $(S,\leq)$.

Let $\CL$ be a class (collection) of posets. A poset
$(S,\leq)\in\CL$ will be called \defn{stem-universal} in $\CL$, if
each poset $(C,\leq)\in\CL$ is isomorphic to some stem in
$(S,\leq)$, i.e. there exists a stem-embedding
$f:(C,\leq)\to(S,\leq)$.
 We call $(S,\leq)$ \defn{homogeneous}, if each isomorphism
$f:(A,\leq)\to(B,\leq)$ between two finite stems of $(S,\leq)$
extends to an automorphism of $(S,\leq)$; equivalently, for any
finite poset $(A,\leq)$ and any two stem-embeddings
$f,f':(A,\leq)\to(S,\leq)$ there exists an automorphism $g$ of
$(S,\leq)$ such that $f'=g\circ f$ (We mention that in the literature there are other concepts called homogeneity. Our homogeneity, for instance, does \emph{not} mean that $\Aut(S)$, the automorphism group of $(S,\leq)$, acts transitively on $S$.).

We say that $(S,\leq)$
\defn{realizes all one-point stem-extensions of finite stems of
$S$}, if whenever $(A,\leq),(B,\leq)$ are finite posets such that
$(A,\leq)\stem(S,\leq)$,$(A,\leq)\stem(B,\leq)$ and $|B|=|A|+1$, then
there exists a stem-embedding $g:(B,\leq)\to(S,\leq)$ such that
$g|_A=\id_A$, the identity on $A$ (this condition includes the case
$A=\emptyset$).

The first main goal of this paper will be the following result

\begin{theorem}\label{thm:unihom}
    Let $\CL$ be the class of all past-finite countable causal sets.
    \begin{enumerate}[(a)]
        \item A poset $(U,\leq)\in\CL$ is stem-universal and
        homogeneous in $\CL$ iff $(U,\leq)$ realizes all one-point
        stem-extensions of finite stems of $U$.
        \item There exists a stem-universal homogeneous past-finite
        causet $(U,\leq)$ in $\CL$. Moreover, $(U,\leq)$ is unique up
        to isomorphism with these properties.
    \end{enumerate}
\end{theorem}

One approach to the proof of Theorem \ref{thm:unihom} would be to use a
category-theoretic generalization of the
Fra\"{\i}ss\'{e}-J\'{o}nsson theorem from model theory, see \cite{DG93},
and to argue directly for past-finite causal sets and stem embeddings.
In order to avoid the category-theoretic machinery needed for this,
here we will base our argument on a more classical version of the
Fra\"{\i}ss\'{e}-J\'{o}nsson theorem, cf. \cite{Fra54,BS69,H93}, which describes
the existence of universal homogenous objects for various classes of
e.g. relational structures. This also has the advantage of showing
how to regard past-finite causal sets and stem-embeddings as relational
structures and relational embeddings. For the convenience of the
reader, we recall these notions. Let $\sigma=(n_i)_{i\in I}$ be a
fixed $I$-indexed sequence of natural numbers $n_i\geq 1$ (a
relational signature) denoting prescribed arities of relations. A
tuple $\A=(A,(R_i)_{i\in I})$ is called a
\defn{$\sigma$-structure}, if $A$ is a set (possibly empty) and
$R_i$ is an $n_i$-ary relation on $A$, i.e. $R_i\subseteq A^{n_i}$,
for each $i\in I$. Given two $\sigma$-structures $\A=(A,(R_i)_{i\in
I})$ and $\B=(B,(Q_i)_{i\in I})$, a mapping $f:A\to B$ is called an
\defn{embedding} of $\A$ into $\B$, if $f$ is one-to-one and for
each $i\in I$ and $x_1,\dots,x_{n_i}\in A$ we have
$(x_1,\dots,x_{n_i})\in R_i$ iff $(f(x_1),\dots,f(x_{n_i}))\in Q_i$.
An embedding which is onto is called an \defn{isomorphism}. An
isomorphism of $\A$ onto itself is called an \defn{automorphism of $\A$}.
Further $\A$ is said to be a \defn{substructure} of $\B$, if
$A\subseteq B$ and $R_i=Q_i\cap A^{n_i}$ for each $i\in I$, i.e.
$A\subseteq B$ and the identity mapping $\id:A\to B$ is an embedding
of $\A$ into $\B$; this is denoted by $\A\subseteq\B$.

Now let $\CL$ be a class (collection) of $\sigma$-structures. A
structure $\U\in\CL$ is called \defn{universal} in $\CL$, if for
each $\A\in\CL$ there exists an embedding $f:\A\to\U$.
Further, $\U$ is \defn{homogeneous}, if each isomorphism $f:\A\to\B$
between two finite substructures $\A,\B$ of $\U$ with $\A,\B\in\CL$
extends to an
automorphism of $\U$. The structure $\U$ \defn{realizes all finite
extensions of finite substructures}, if whenever $\A,\B\in\CL$ are
finite structures such that $\A\subseteq\U$ and $\A\subseteq\B$,
then there exists an embedding $g:\B\to\U$ such that $g|_A=\id_A$.
The class $\CL$ is said to be an \defn{$\omega$-class}, if it
satisfies the following conditions:

\begin{enumerate}[(1)]
    \setcounter{enumi}{-1}
    \item Any $\A\in\CL$ is countable.
    \item Whenever $\A\in\CL$ and $\B$ is a structure isomorphic to
    $\A$, then $\B\in\CL$.
    \item If $\A_1\subseteq\A_2\subseteq\dots$ are finite structures
    from $\CL$, forming a chain of substructures under inclusion, then
    their union $\A=\bcup_{i\in\N}\A_i$ (whose domain and relations
    are defined as the union of the domain resp. corresponding
    relations of the structures $\A_i$) also belongs to $\CL$.
    \item If $\A\in\CL$ and $F$ is a finite subset of the domain of
    $\A$,
    then there exists a finite substructure $\S\subseteq\A$ with
    $\S\in\CL$ whose domain contains $F$.
\end{enumerate}

An object $\A\in\CL$ is called \defn{weakly initial in $\CL$}, if
for each $\B\in\CL$ there exists an embedding $f:\A\to\B$. (Often,
this is the empty or a singleton structure.) The class $\CL$ is said
to have

\begin{itemize}
    \item the \defn{joint embedding property}, if for any $\A,\B$ there
    exists $\C\in\CL$ and embeddings $f:\A\to\C,g:\B\to\C$
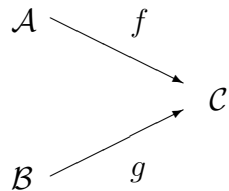
    \begin{figure}[h]
        \begin{picture}(100,100)(0,-20)
            \put(230,0){$g$}
            \put(230,55){$f$}
            \put(185,55){$\A$}
            \put(185,-5){$\B$}
            \put(260,25){$\C$}
            \put(200,0){\vector(2,1){50}}
            \put(200,60){\vector(2,-1){50}}
        \end{picture}
    \end{figure}
    \item the \defn{amalgamation property}, if for any
    $\A,\B_1,\B_2\in\CL$ and embeddings $f_i:\A\to\B_i$ ($i=1,2$)
    there exists $\C\in\CL$ and embeddings $g_i:\B_i\to\C$ ($i=1,2$)
    such that $g_1\circ f_1=g_2\circ f_2$, i.e. the subsequent
    diagram ''commutes''.
    \begin{figure}[h]
        \begin{picture}(100,100)(-20,-20)
            \put(115,25){$\A$}
            \put(185,55){$\B_1$}
            \put(185,-5){$\B_2$}
            \put(260,25){$\C$}
            \put(150,55){$f_1$}
            \put(150,-5){$f_2$}
            \put(225,55){$g_1$}
            \put(225,-5){$g_2$}
            \put(200,0){\vector(2,1){50}}
            \put(200,60){\vector(2,-1){50}}
            \put(130,35){\vector(2,1){50}}
            \put(130,25){\vector(2,-1){50}}
        \end{picture}
    \end{figure}
\end{itemize}

\begin{theorem}[\cite{BS69,Fra54,H93}]\label{thm:fra}
Let $\sigma$ be a relational signature and $\CL$ an $\omega$-class
of $\sigma$-structures.
    \begin{enumerate}[(a)]
        \item Let $\CL$ contain a weakly initial structure. Then a
        structure $\U\in\CL$ is universal and homogeneous in $\CL$ if
        and only if $\U$ realizes all finite extensions of finite
        substructures.
        \item The following are equivalent:
        \begin{enumerate}[(1)]
            \item $\CL$ contains a universal homogeneous structure
            $\U$.
            \item $\CLfin$, the class of all finite structures in
            $\CL$, has the joint embedding and the amalgamation
            property and contains up to isomorphism only countably
            many structures.
        \end{enumerate}
        Moreover, in this case $\U$ is unique up to isomorphism.
    \end{enumerate}
\end{theorem}


We just note that the proof of Theorem \ref{thm:fra}(a) as well as of 
the uniqueness of $\U$ in part (b) employs a standard `back-and-forth 
argument'. The construction of the universal homogeneous structure $\U$ 
given condition (2) of part (b) uses a suitable enumeration of all 
possible embeddings of the finite structures in $\CL$.


In order to be able to apply Theorem \ref{thm:fra} to causal sets,
we have to enrich them to relational structures such that
stem-embeddings become relational embeddings as above. Given a poset
$(A,\leq)$, we define its \defn{relational expansion} to be
$(A,\leq,\R)$ where $\R=(R_i)_{i\geq 1}$ and $R_i=\{x\in A\mid
|\past(x)|=i\}$ ($i\geq1$). Hence $\leq$ is a binary and each $R_i$
a unary relation on $A$, so $(A,\leq,\R)$ is a $\sigma$-structure
for the signature $\sigma=(2,1,1,1,\dots)$. Observe that if $A$ is
finite, then each of the relations $R_i$ where $i>|A|$ is empty.
Next we show that for past-finite causal sets, this expansion achieves
our first goal:

\begin{proposition}
\label{prop}
Let $(A,\leq)$ and $(B,\leq)$ be two past-finite causal sets and let
$(A,\leq,\R)$ resp. $(B,\leq,\Q)$ be their relational expansions.
\begin{enumerate}[(a)]
    \item $(A,\leq)$ is a stem of $(B,\leq)$ iff $(A,\leq,\R)$ is a
    substructure of $(B,\leq,\Q)$.
    \item Let $f:A\to B$ be a mapping. Then $f:(A,\leq)\to(B,\leq)$
    is a stem-embedding iff $f:(A,\leq,\R)\to(B,\leq,\Q)$ is an
    embedding of relational structures.
\end{enumerate}

\begin{proof}
(a) We have $\R=(R_i)_{i\geq1}$ and $\Q=(Q_i)_{i\geq1}$ with
$R_i=\{x\in A\mid|\past_A(x)|=i\}$ and $Q_i=\{y\in B\mid
|\past_B(y)|=i\}$ ($i\geq 1$).

First, let $A\stem B$. Let $x\in A$. By $A\stem B$ we get
$\past_A(x)=\past_B(x)$, so $x\in R_i$ iff $x\in Q_i$, for each
$i\geq1$. Hence $(A,\leq,\R)\subseteq(B,\leq,\Q)$. Conversely,
assume the latter. Let $x\in A$ and $y\in B$ with $y\leq x$. By
assumption, we have $\past_A(x)\subseteq\past_B(x)$ and
$|\past_A(x)|=|\past_B(x)|\in\N$, so $\past_A(x)=\past_B(x)\ni y$
showing $y\in A$. Hence $A\stem B$.

\bs (b) Straightforward by (a), using that the image of $A$ under
$f$ is an isomorphic copy of $A$ and a stem resp. a substructure of
$B$.
\end{proof}
\end{proposition}

This result allows us to translate all notions concerning
stem-embeddings (like universality, homogeneity,\dots) into
corresponding ones for the relational expansions and their
embeddings.

Let $\CL$ be the class of all countable past-finite causal sets and let
$\CLrel$ be the collection of all relational expansions
$(A,\leq,\R)$ where $(A,\leq)\in\CL$. Next we give an easy
application of Proposition \ref{prop} to show how this translation of
notions works:

\begin{remark}
\label{rem}
$\CLrel$ is an $\omega$-class of $\sigma$-structures.
\end{remark}

\begin{proof}
Conditions (0) and (1) of the definition of $\omega$-class are
trivial. For (2), let\\
$(A_1,\leq,\R_1)\subseteq(A_2,\leq,\R_2)\subseteq\dots$ be a chain
of finite structures from $\CLrel$, and let $(A,\leq,\R')$ be their
union. Then $(A_1,\leq)\stem(A_2,\leq)\stem\dots$ is a sequence of
stems by Propositions \ref{prop} (a), and $(A,\leq)$ is a past-finite
causal set whose expansion $(A,\leq,\R)$ coincides with
$(A,\leq,\R')$. Hence $(A,\leq,\R')\in\CLrel$. To check (3), let
$(A,\leq,\R)\in\CLrel$ and let $F$ be a finite subset of $A$. Then
$\past(F)$ is a finite stem of $(A,\leq)$, so by Proposition \ref{prop}
(a) its relational expansion is a finite substructure of
$(A,\leq,\R)$ belonging to $\CLrel$.
\end{proof}

Next we show:

\begin{lemma}
\label{lem}
Let $(A,\leq),(B,\leq)$ be two finite posets such that $A\stem B$.
Then there is a sequence of stems
\[A=A_0\stem A_1\stem\dots\stem A_m=B\]
such that $|A_{i+1}|=|A_i|+1$ for each $i=0,\dots,m-1$.
\end{lemma}

\begin{proof}
By induction on $|B\setminus A|$. 
Choose a minimal element $x$ of $B\setminus A$, and put 
$A_1=A\cup\{x\}$. Then $A\stem A_1\stem B$, and by
induction we
obtain a sequence of stems and one-point extensions from $A_1$ to $B$.
\end{proof}

Now we can give the

\begin{proof}[Proof of Theorem \ref{thm:unihom}]
By Remark \ref{rem}, $\CLrel$ is an $\omega$-class of
$\sigma$-structures.

\bs (a) The empty structure is a weakly initial object of $\CLrel$.
Lemma \ref{lem} shows that if $\U$
realizes all one-point extensions of finite stems, it also realizes
all finite stem-extensions of finite stems. Now the result is a
translation of Theorem \ref{thm:fra} (a).

\bs (b) Again we use Proposition \ref{prop}. We check condition (2)
of Theorem \ref{thm:fra} (b). It is clear that $\CLrel$ contains up
to isomorphism only countably many finite structures $(A,\leq,\R)$,
since if $A$ is finite, only finitely many of the relations $R_i$
are non-empty. It remains to show that the collection of finite
posets satisfies the joint embedding and the amalgamation properties
with respect to stem-embeddings. Since this collection contains the
empty poset, it suffices to check the amalgamation property. For
this, let $(A_i,\leq_i)$ ($i=0,1,2$) be three finite posets such
that $(A_0,\leq_0)\stem(A_i,\leq_i)$ for $i=1,2$. We may assume that
$A_0=A_1\cap A_2$. Put $A=A_1\cup A_2$ and $\leq=\leq_1\cup\leq_2$,
i.e. for $x,y\in A$ let $x\leq y$ iff either $x,y\in A_1$ and
$x\leq_1 y$ or $x,y\in A_2$ and $x\leq_2 y$. Then $\leq$ is
transitive, since if e.g. $x\leq_1 y$ and $y\leq_2 z$, we have $y\in
A_1\cap A_2=A_0$ and $x\in A_1$. Since $A_0\stem A_1$, we get $x\in
A_0$ and $x\leq_0 y$, so $x\leq_2 y\leq_2 z$ which implies $x\leq
z$. Hence $\leq$ is a partial order on $A$. Observe that if
$x\in A_1\setminus A_0$ and $y\in A_2\setminus A_0$, then neither $x\leq y$
nor $y\leq x$. Also note that if $x,y\in
A_1\cap A_2$ and $x\leq_1 y$, say, then this implies $x\leq_0 y$,
hence also $x\leq_2 y$.
We claim that
$(A_i,\leq_i)\stem(A,\leq)$ for $i=1,2$. By the remark just made,
$\leq_i$ is just the restriction of $\leq$ to $A_i$. So, it only
remains to show that $A_i$ is a stem of $(A,\leq)$.

Indeed, let $x\in A$ and $y\in A_1$ with $x\leq y$. We claim that
$x\in A_1$. This is trivial if $x\leq_1 y$. So let $x\leq_2 y$. Then
$x,y\in A_2$ and $y\in A_0$. Now $A_0$ is a stem of $(A_2,\leq)$, so
$x\in A_0\subseteq A_1$. Hence $A_1$ is a stem of $(A,\leq)$ and for
$A_2$ we argue analogously.
 This proves the amalgamation property.

Now the result follows from Theorem \ref{thm:fra} (b).
\end{proof}

Next we wish to describe the structure of the universal homogeneous past-finite causet $(U,\leq)$ further. A poset $S,\leq)$ is called \defn{directed}, if for any $a,b\in S$ there is $c\in S$ with $a\leq c$ and $b\leq c$. Two elements $a,b\in S$ are \defn{incomparable}, if neither $a\leq b$ nor $b\leq a$; this is denoted by $a\|b$. A subset $A\subseteq S$ is called an \defn{antichain}, if any two  elements of $A$ are incomparable. An element $x\in S$ is called \defn{maximal}, if there is no $y\in S$ with $x<y$. Together with Theorem \ref{thm:unihom}(a), the following provides an order-theoretic characterization of the structure of the universal homogeneous past-finite causal set. 
\begin{proposition}\label{prop:realize}
Let $(U,\leq)$ be a past-finite causal set. The following are equivalent;
\begin{enumerate}[(1)]
\item $(U,\leq)$ realizes all one-point stem-extensions of finite 
stems of $U$.
\item \begin{enumerate}[(i)]
\item $(U,\leq)$ is directed, and 
\item For any finite antichain $A\subseteq U$ (including the case 
that $A=\emptyset$) and any $y\in U\setminus A$ with $A\subseteq\past(y)$ 
there is $x\in U\setminus A$ such that $x\|y$ and $\past(x)=\past(A)\cup\{x\}$. 
\end{enumerate}
\end{enumerate}
In this case, no element of $(U,\leq)$ is maximal.
\end{proposition}
Note that the condition $\past(x)=\past(A)\cup\{x\}$ means that $a\leq x$ for each $a\in A$, and whenever $u\in U$ with $u<x$, then $u<a$ for some $a\in A$.
\begin{proof}
\underline{$(1)\rightarrow (2)$}; To show (i), let $a,b\in U$. We claim that there is $c\in U$ with $a\leq c$ and $b\leq c$. Clearly, we may assume that $a\|b$. Let $A=\past(\{a,b\})$, choose an element $z\notin A$, and put $B=A\cup\{z\}$. We define a partial order $\leq^*$ on $B$ such that it extends the order $\leq$ on $A$ and $a\leq^* z,b\leq^* z$, so $z$ is the greatest element of $(B,\ls)$. Then $(A,\leq)\stem(B,\ls)$, a one-point extension of the finite stem $A$ of $U$. Hence, there exists a stem-embedding $g: (B,\ls)\to (U,\leq)$ with 
$g|_{A}=\id_A$. Now put $c=g(z)$ to obtain the chain. 

For (ii), choose a finite antichain $A\subseteq U$ and $u\in U\setminus A$ with $A\subseteq \past_U(y)$. Put $A'=\past_U(y)$, choose an element $z\notin A'$, and put $B=A'\cup\{z\}$. 
Define a partial order $\ls$ on $B$ such that it extends the order 
$\leq$ on $A$, such that each $x\in\past_U(A)$ satisfies $x\ls z$, 
but each $x\in A'\setminus\past_U(A)$ is incomparable to $z$ in $(B,\ls)$.
Then $(A',\leq)\stem(B,\ls)$, a one-point extension of the finite stem $A'$ of $U$. Again, there is a stem-embedding 
$g: (B,\ls)\to (U,\leq)$ with $g|_{A'}=\id_{A'}$. Then $x=g(z)$ satisfies $x\notin A'$ and $x\|z$, and by $g(B)\stem U$ we obtain 
\begin{eqnarray*}
\past_U(x)&=&\past_{g(B)}(x)=g(\past_B(z))=g(\past_B(A)\cup\{z\}) \\
  &=& \past_{g(B)}(A)\cup\{x\}=\past_U(A)\cup\{x\},
\end{eqnarray*}
as required.

\bs \underline{$(2)\rightarrow (1)$}; We first show that no element of $(U,\leq)$ is maximal. Let $y\in U$. Applying assumption (ii) to the empty antichain, we obtain $x\in U$ with $x\|y$. Now by (i) there is $z\in U$ with $x\leq z$ and $y\leq z$. Then $y<z$.

Now let $(A'\leq)$ and $(B,\leq')$ be finite posets such that $(A'\leq)\stem (U,\leq),(A'\leq)\stem(B,\leq')$, and $B=A'\cup\{z\}$, say. Let $A$ comprise all elements $a\in A'$ maximal with respect to $a<'z$ in $(B,\leq')$. So, $A$ is an antichain and $\past_B(z)=\past_B(A)\cup\{z\}$. Since $(U,\leq)$ is directed
and $A'$ is finite, there exists $y\in U$ with $a\leq y$ for each $a\in A'$. Choose $y'\in U$ with $y<y'$; then $y'\notin A'$. Now by assumption (ii), there is $x\in U\setminus A$ such that in $(U,\leq)$ we have $x\|y'$ and $\past_U(x)=\past_U(A)\cup\{x\}$. Define $g:B\to U$ by $g|_{A'}=\id_{A'}$ and 
$g(z)=x$. By $x\|y'$, we obtain $x\notin A'$, and moreover, $x$ is incomparable to each element of $A'\setminus \past_U(A)$. Hence $g$ is an embedding, and we show that $g(B)\stem U$. We have $A'\stem U$, and so
$\past_U(x)=\past_U(A)\cup\{x\}\subseteq A'\cup\{x\}=g(B)$.
Thus $g$ is a stem-embedding.
\end{proof}
Next we wish to use Theorem \ref{thm:fra}(a) and Proposition 
\ref{prop:realize} to give a direct construction (avoiding 
Theorem \ref{thm:fra}(b)) of the universal homogeneous past-finite causal set $(U,\leq)$. We are thankful to an anonymous referee who pointed out this alternate construction. 
\begin{construction}[of the universal homogeneous past-finite causal set $(U,\leq)$]\label{con:u}

We construct a sequence of finite posets $(A_1,\leq_1)\stem(A_2,\leq_2)\stem\ldots$ as follows. Let $(A_1,\leq_1)$ be any singleton set, with the trivial order. Now let $i\in\N$ and assume we have constructed a finite poset $(A_i,\leq_i)$. Let $\A$ comprise all antichains in $(A_i,\leq_i)$, and choose pairwise different elements $x_A\  (A\in\A)$ not belonging to $A_i$. Then put $A_{i+1}=
A_i\cup\{x_A\mid A\in\A\}$, and define a partial order $\leq_{i+1}$ on $A_{i+1}$ such that it extends $\leq_i$ and for any $z\in A_{i+1}$, if $z\in\past_{A_i}(A)\cup\{x_A\}$ then $z\leq_{i+1} x_A$, but if $z\notin \past_{A_i}(A)\cup\{x_A\}$ then $z$ and $x_A$ are incomparable in $(A_{i+1},\leq_{i+1})$. Hence each element $x_A\  (A\in\A)$ is maximal in $(A_{i+1},\leq_{i+1})$, and thus $(A_i,\leq_i)\stem (A_{i+1},\leq_{i+1})$. Finally, let $(U,\leq)=\bigcup_{i\in\N}(A_{i+1},\leq_{i+1})$. Clearly, 
$(A_i,\leq_i)\stem (U,\leq)$ for each $i\in\N$, and $(U,\leq)$ is a countable past-finite causal set.

\bs \underline{Claim:} $(U,\leq)$ is universal and homogeneous.

\end{construction}
\begin{proof}
We show that $(U,\leq)$ satisfies condition (2) of Proposition \ref{prop:realize}; then this proposition and Theorem \ref{thm:unihom}(a) imply the result.
For condition (2)(i), let $a,b\in U$ with $a\|b$. Then $A=\{a,b\}\subseteq A_i$ for some $i\in\N$, and $x_A\in A_{i+1}$ satisfies $a\leq x_A$ and $b\leq x_A$.  For condition (2)(ii), let $A\subset U$ be a finite antichain and $y\in U\setminus A$ with $A\subseteq \past_U(y)$. Again, $A\cup\{y\}\subseteq A_i$ for some $i\in\N$. Then $x_A\in A_{i+1}$ and by $A_{i+1}\stem U$ and construction of $\leq_{i+1}$ we obtain
$$\past_U(x_A)=\past_{A_{i+1}}(x_A)=\past_{A_{i+1}}(A)\cup\{x_A\}=\past_U(A)\cup\{x_A\}.$$
Furthermore, since $y\notin \past_{A_{i+1}}(A)$, we get $x_A\| y$ in $(A_{i+1},\leq_{i+1})$ and in $(U,\leq)$, as required. Our claim follows.
\end{proof}
This argument is more direct and intuitive than the previous one using Theorem \ref{thm:fra}(b) with its proof. However, Theorem \ref{thm:fra} puts the result into a general context, and checking the amalgamation property required for condition (2) of Theorem \ref{thm:fra}(b) was also uncomplicated.

Next we mention two further structural properties of the universal homogeneous past-finite causet $(U,\leq)$. 


\begin{corollary}\label{cor:cover}
Let $(U,\leq)$ be the universal homogeneous countable past-finite
causet.
\begin{enumerate}[(a)]
    \item For any finite subset $F\subseteq U$, $(U,\leq)$ is
    isomorphic to the poset $(U\setminus F,\leq)$. In particular,
    $(U,\leq)\cong(U\setminus\past(z),\leq)$ for each $z\in U$.
    \item For each $z\in U$, $(U,\leq)$ is isomorphic to $(\{u\in
    U\mid z<u\},\leq)$, the proper future of $z$.
\end{enumerate}
\end{corollary}

\begin{proof}
(a) We may assume that $F\neq\emptyset$. Trivially, $(U\setminus
F,\leq)$ is a countable past-finite causet. 
We wish to check that $(U\setminus F,\leq)$ satisfies condition (2) of Proposition \ref{prop:realize}; then the result is immediate by Proposition \ref{prop:realize} and Theorem \ref{thm:unihom}. Since $(U,\leq)$ is directed and contains no maximal element,  $(U\setminus F,\leq)$ is directed. So, let $A\subseteq 
U\setminus F$ be a finite antichain and $y\in (U\setminus F)\setminus A$ with 
$A\subseteq \past_{U\setminus F}(y)$. Again, since $(U,\leq)$ is directed and contains no maximal element, there is $y'\in U\setminus F$ with $F\cup\{y\}\subseteq \past_U(y')$. By Proposition \ref{prop:realize}, for $(U,\leq)$, there is
$x\in U\setminus A$ such that 
$x\| y'$ and $\past_U(x)=\past_U(A)\cup\{z\}$. Then $x\notin F$ and neither 
$x\leq y$ nor $y\leq x$, and clearly
$\past_{U\setminus F}(x)=\past_{U\setminus F}(A)\cup\{x\}$, as required for condition (ii).
The final claim is immediate, since
$\past(z)$ is finite for each $z\in U$.

\bs (b) Let $z\in U$ and $U'=\{u\in U\mid z<u\}$. Again we show that
$(U',\leq)$ satisfies condition (2) of Proposition \ref{prop:realize}.
If the antichain $A$ chosen is empty, replace it by $\{z\}$. Then, and also in case $A$ is non-empty, apply condition (ii) for $(U,\leq)$ to obtain the element $x\in U'$ as required.
%
\end{proof}

Finally, we wish to show that the restriction to \emph{past-finite}
causets in Theorem \ref{thm:unihom} is essential:

\begin{proposition}\label{prop:prop2}
The class of all countable causets does \emph{not} contain a universal
causet.
\end{proposition}

\begin{proof}
We will exploit that this class also contains posets which are not 
past-finite.
Suppose there was a countable causet $(U,\leq)$ such that each
countable causet can be stem-embedded into $(U,\leq)$.
Let $-\N$ denote the set of negative integers. We denote the natural partial order on $-\N$ by $\preceq$, so $n-1\prec n$ for each $n\in -\N$.
For each subset
$A\subseteq -\N$ let $(S_A,\leq)$ be the causet obtained from $(-\N,\preceq)$
by replacing each element $n\in A$ by a 2-element antichain $\{n,n^*\}$.
That is, $S_A=-\N\cup\{n^*\mid n\in A\}$, $n$ and $n^*$ are incomparable
for each $n\in A$, and $m<n^*$ iff $m<n$ iff $m\prec n$
($m\in -\N,n\in A$), likewise for $m^*<n^*$ resp. $m^*<n$. By assumption,
there is a stem-embedding $f_A:(S_A,\leq)\to(U,\leq)$. Let $x_A=f_A(-1)$.
Then $f_A(S_A)$ is a stem of $U$, and $f_A(S_A)=\past(x_A)$.

As is easily seen, for $A,A'\subseteq -\N$ we have $(S_A,\leq)\cong(S_{A'},\leq)$
iff $A=A'$. Hence there are uncountably many non-isomorphic causets
of the form $(S_A,\leq)$; these are isomorphic to the posets $(\past(x_A),\leq)$.
But $U$ is countable and contains only countably many posets of the form $\past(x)$
($x\in U$), a contradiction.
\end{proof}

\section{Probabilistic constructions}

In this section, we wish to provide probabilistic constructions of
the universal homogeneous causal set $(U,\leq)$ of section $2$.
Probabilistic constructions of causal sets were already investigated
in \cite{G03,RS00}. This employed successive one-point extensions
(``generalized percolations'') of finite posets; such extensions will also be
used here. Suppose we are given a finite poset $(A,\leq)$ and want
to extend it to a poset $(B,\leq)$ such that $B=A\cup\{z\}$, say,
and $(A,\leq)$ is a stem in $(B,\leq)$. Then we just have to define
the order relations between $z$ and the elements of $A$; moreover,
since $A$ should become a stem of $B$, $z$ has to become a maximal
element of $(B,\leq)$, thus for each $a\in A$ we can only put $a<z$
or $a\|z$, and this can be decided probabilistically. 
Let us now give the
details of this construction for our case. We will discuss a 
technical difference to \cite{G03,RS00} afterwards. 
For sake of concreteness, we will take the underlying set of our causet to be $\N$
(as in \cite{RS00}, but the construction would work for any countably
enumerated set). Hence we will construct a partial order $\leq$ on $\N$. Recall that we write $x<y$ if $x\leq y$ and $x\neq y$. We denote the natural order on $\N$ by $\preceq$, i.e. $n\prec n+1$ for each $n\in\N$.
Our
basic construction will depend on a parameter $p\in[0,1]$.

\begin{construction}[Probabilistic one-point
extension of an enumerated finite poset]\label{con1} Let
$p\in[0,1]$. Let $(A,\leq)$ be a finite poset with
$A=\{1,\dots,n\}$ for some $n\succeq 1$, and let $B=A\cup\{n+1\}$.
Furthermore, letting $A_j=\{1,\dots,j\}$ for $1\preceq j\preceq n$, assume
that $A_1\stem\dots\stem A_n=A$. As noted before,
we wish to extend the order $\leq$ to a partial order $\leq'$ on $B$
such that $(A,\leq)$ is a stem of $(B,\leq')$. 
We proceed as follows. First choose, with equal probability, some $j\in\{1,\ldots,n\}$. If $j$ was chosen, this means that we will restrict ourselves to having
$\past_B(n+1)\subseteq A_j\cup\{n+1\}$. We define a binary relation $R$ in 
$\{1,\ldots,n\}\times\{n+1\}$ as follows. Decide
independently for each $i\in \{1,\ldots,j\}$ with probability $p$ that $(i,n+1)\in R$, and with probability $1-p$ that $(i,n+1)\notin R$. Now let $\leq'$ on $B$ be the reflexive transitive closure of the relation $\leq\cup R$. That is, we have $n+1\leq' n+1$, for $x,y\in A$ we have 
$x\leq' y$ iff $x\leq y$, and for any $x\in A$ we have $x\leq' n+1$ iff there is
$i\in \{1,\ldots,j\}$ such that $x\leq i$ and $(i,n+1)\in R$; in this case $x\in A_j$ by $A_j\stem A$. Clearly, $(B,\leq')$ is a partial order, $n+1$ is maximal in $(B,\leq')$, $\past_B(n+1)\subseteq A_j\cup\{n+1\}$, and $(A,\leq)\stem (B,\leq')$.
\end{construction}

\bs Next we wish to construct our random past-finite countable causal
set structure on $\N$.

\begin{construction}[of a probabilistic order $\leq$ on
$\N$]\label{con2} Let $p\in[0,1]$. Let $A_n=\{1,2,\dots,n\}$
($n\in\N$). For $n=1$, put $1\leq1$. Now use Construction \ref{con1}
to successively extend the order $\leq$ from $A_n$ to $A_{n+1}$. We
obtain a sequence of stems $A_1\stem A_2\stem\dots\stem
A_n\stem\dots$, and we put $(\N,\leq)=\bcup_{n\in\N}(A_n,\leq)$.
Then each $A_n$ is a stem of $(\N,\leq)$ and $(\N,\leq)$ is a
past-finite causal set.
\end{construction}

\bs Next we will show:

\begin{theorem}
\label{thm:prob} Let $p\in(0,1)$. With probability 1, the above
construction produces a partial order $\leq$ on $\N$ such that
$(\N,\leq)$ is a stem-universal homogeneous past-finite causal set.
\end{theorem}

\begin{proof}
As noted before, $(\N,\leq)$ is a past-finite causal set. By Theorem
\ref{thm:unihom} (a), it suffices to show that with probability 1, $(\N,\leq)$
realizes all one-point stem-extensions of finite stems of
$(\N,\leq)$. There are countably many such stem-extensions. Since
the intersection of countably many events of probability 1 again has
probability 1, it suffices to consider an arbitrary fixed one-point
stem-extension $A\stem(B,\leq')$ with $B=A\cup\{y\}$, $A$ finite and
$A\stem(\N,\leq)$. We claim that with probability 1 there exists
$z\in\N$ such that the mapping $g:B\to\N$ with $g|_A=\id_A$ and
$g(y)=z$ is a stem-embedding, i.e., for each $a\in A$ we have
$a\leq' y$ in $B$ iff $a\leq z$ in $\N$ and
$A\cup\{z\}\stem(\N,\leq)$.

For each $j\in\N$ let $A_j=\{1,2,\dots,j\}$. Since $A$ is finite,
there is $m\in\N$ such that $A\subseteq A_m$. Then $A\stem A_m$. Now
choose any integer $n\succeq m$ and consider the extension of the order $\leq$
from $A_n$ to $A_{n+1}$ given by Construction \ref{con1}. We 
wish to compute a lower bound for the probability that we can put $z=n+1$ to obtain the required stem $A\cup\{z\}\stem (\N,\leq)$.
First, the probability that we choose the number $m$ from $\{1,\dots,n\}$ is $1\over n$.
Then by Construction \ref{con1} for each $i\in A$, if $i<'y$ in $(B,\leq')$ we put $(i,n+1)\in R$
with probability $p$, and if $i||'y$ we put $(i,n+1)\not\in R$ with probability $1-p$;
further with probability $1-p$ we put $(i,n+1)\notin R$ for each $i\in A_m\setminus A$. Hence, given $m$, there is a (small but) fixed $r>0$, depending only on the structure of $(A_m,\leq)$, on $A\stem B$ and on $y$ (but not on $n$), such that 
at least with probability $r$
all $i\in A_m$ satisfy the above conditions, so $A\cup\{n+1\}\stem A_m\cup\{n+1\}\stem A_{n+1}$ and $n+1$ realizes the one-point stem extension of $A$ as required. 

Thus the probability that in $(A_{n+1},\leq)$, the structure of $(A_m\cup\{n+1\},\leq)$ is as required is at least $r\cdot {1\over n}$. Hence the probability that $n+1$ and $(A_m\cup\{n+1\},\leq)$ do not satisfy these conditions is at most $1-{r\over n}$. 
Consequently, the probability that \emph{no} integer $n\succeq m,n+1$ and 
$(A_m\cup\{n+1\})$ realizes this one-point extension of $A$ is 
$\prod_{n=m}^{\infty}(1-{r\over n})$. 
Since the series $\sum_{n=m}^{\infty}{r\over n}$ diverges, we obtain 
$\prod_{n=m}^{\infty}(1-{r\over n})=0$. 
Hence with probability 1, this
one-point extension of $A$ is realized inside $(\N,\leq)$ as
required, and the result follows.
\end{proof}

We just note that the above `intuitive' probabilistic statements
can easily be made exact by a precise (but technical) definition of the
employed probability space; this can be done analogously to the case
of the random graph, cf. \cite{ER63,H93}, or to the procedure in
\cite{RS00,BDGHS01,BDGHS03} for constructing the probability space
of completed labelled causets: The sample space of this probability space is the collection
of all past-finite causets $(\N,\leq)$
(hence the underlying set, $\N$, is fixed, but the order $\leq$ can vary), 
and Construction \ref{con1}
yields a corresponding probability measure. Theorem \ref{thm:prob}
can then be rephrased by saying that in this probability space the
collection of all causets which are universal and homogeneous has measure $1$.

The above construction would work also for other probability distributions on $\{1,\ldots,n\}$, in Construction \ref{con1}, than the uniform one. However, they would have to be chosen with some care in order to ensure the final argument in the proof of Theorem \ref{thm:prob}. 

Our construction of $(U,\leq)$ uses and depends
on the given enumeration of $\N$, resp., in Construction \ref{con1},
of the poset $A=\{1,\dots,n\}$, and not only on the structure of the
poset $(A,\leq)$:
After choosing $j\in\{1,\ldots, n\}$, we decided to possibly put $(i,n+1)\in R$ only for elements $i\in\{1,\ldots,j\}$.
This is a bit unfortunate, since, in the notions of
\cite{RS00}, it would correspond to some externally given time. 
Apart from this, our construction is very similar to the ones proposed in \cite{G03,RS00}. It even shares with it a property of the ``gregarious child transition'', cf. \cite[Lemma 2]{RS00}:
\begin{remark}\label{card}
Let $p\in[0,1]$, let $(A,\leq)$ be a finite poset with $A=\{1,\ldots,n\}$, and let 
$B=A\cup\{n+1\}$. The probability to obtain by Construction \ref{con1} the poset 
$(B,\leq)$ with $(A,\leq)\stem (B,\leq)$ and $n+1\| a$ for each $a\in A$ depends only on the cardinality of $A$. 
\end{remark}
\begin{proof}
Given $j\in \{1,\ldots,n\}$, for each $i\in \{1,\ldots,j\}$ we have to put $(i,n+1)\notin R$. The probability of this equals ${1\over  n}\cdot \sum_{j=1}^n (1-p)^j$.
\end{proof}
Furthermore, our construction apparently satisfies a weak form of the Bell causality condition, but not the condition of discrete general covariance of \cite{G03,RS00}. It would be interesting to investigate this further. Also, this 
 raises the question whether the universal
homogeneous past-finite causet $(U,\leq)$ can be constructed without
referring 
as in Construction \ref{con1} 
to a given enumeration, with positive probability, say.
This would require a more
intricate analysis of the probabilities of the finite posets
occurring.

Finally, we wish to present also a number-theoretic
representation of the universal homogeneous past-finite causet $(U,\leq)$.
As underlying set, we take again the natural numbers $\N$.

\begin{construction}[of a partial order $\leq$ on $\N$]\label{order}
We define a binary relation $R$ on $\N$ as follows. For any $j,n\in\N$, put $(j,n)\in R$ iff 
$j\prec n$ and 
in the unique ternary expansion of $n$ as a sum of distinct powers of $3$, $3^j$ occurs with coefficient $1$. 
That is, $(j,n)\in R$ iff
$$n=3^j+
\sum_{0\preceq i\preceq n\atop i\neq j}
x_i\cdot 3^i 
$$
for some  $x_i\in\{0,1,2\}$. Then let $\leq$ be the transitive reflexive closure of $R$.
\end{construction}
Since $(j,n)\in R$ implies $j\prec n$, clearly $(\N,\leq)$ is a past-finite causal set, for  $A_n=\{1,\ldots,n\}$ we have $A_n\stem \N$, and $n$ is a maximal element of $(A_n,\leq)$. We show:

\begin{theorem}
The poset $(\N,\leq)$ constructed above is a stem-universal homogeneous past-finite causal set.
\end{theorem}

\begin{proof}
By Theorem \ref{thm:unihom} (a), it suffices to show that $(\N,\leq)$ realizes all
one-point extensions. So let $(A,\leq)\stem(\N,\leq)$, with $A$ finite, $(A,\leq)\stem(B,\leq')$
and $B=A\cup\{y\}$, say, with $y\notin A$.
We claim that there exists $z\in\N$ such that the mapping
$g:B\to\N$ with $g|_A=\id_A$ and $g(y)=z$ is a stem-embedding. 
Choose $m\in\N$ such that $A\subseteq A_m$. 
Put
\[z=\sum_{a\in A\atop a<' y}3^a + 
2\cdot 3^m.\]
Then $z\notin A$ and $z$ is maximal in $(A\cup\{z\},\leq)$.
The construction of $\leq$ shows that if $a\in A$ with $a<' y$, then $(a,z)\in R$, so $a<z$. Conversely, 
let $a\in \N$ and $a<z$. Since $\leq$ is the transitive reflexive closure of $R$, there is $a'\in\N$ such that $a\leq a'$ and $(a',z)\in R$.
By definition of $R$ and of $z$, we immediately get $a'\in A$ and $a'<' y$. Since $A\stem\N$, we obtain
$a\in A$ and $a<' y$. 
It follows
that $g$, defined as above, is an order-embedding of $(B,\leq')$ into $(\N,\leq)$
and that $g(B)$ is a stem of $\N$, as required.
\end{proof}

\section{Discussion}
We studied the class of all countable causal sets. In algebra, many classes of structures have been investigated with respect to the existence of universal or homogeneous objects. Most often, this completely depends on the given class of structures. Here, we could show that the class of all countable past-finite causets, in contrast to the class of all causets, contains a 
universal object. Somehow surprizingly, a random (intuitively: `chaotic') construction produces, in the end, almost surely a universal causet bearing maximal degree of symmetry.

A basic idea behind causal set theory is that a manifold $M$ may emerge from a causal set $(S,\leq)$ by some sprinkling of a coarse-grained version of 
$(S,\leq)$ densely into $M$ (cf. \cite{S02}). Then it would be interesting to see how symmetry properties of $(S,\leq)$ are reflected in the structure of $M$. It is tempting to recall, in this context, Noether's close correspondence 
between symmetries and conservation laws. A bold question: Can conservation laws be founded on symmetry properties of causal sets and, ultimately, traced to some random constructions of causal sets?

\nocite{*}\bibliography{uhrcs}

\begin{thebibliography}{10}

\bibitem{AM03}
A.~Ash and P.~McDonald.
\newblock {Moment problems and the causal set approach to quantum gravity}.
\newblock {\em J. Math. Phys.}, 44:1666--1678, 2003, [arXiv:gr-qc/0209020].

\bibitem{BS69}
J.L. Bell and A.B. Slomson.
\newblock {\em {Models and Ultraproducts: An Introduction}}.
\newblock North Holland, Amsterdam, 1969.

\bibitem{BLMS87}
L.~Bombelli, J.~Lee, D.~Meyer, and R.D. Sorkin.
\newblock {Spacetime as a causal set}.
\newblock {\em Phys. Rev. Lett.}, 59:521--524, 1987.

\bibitem{BDGHS03}
G.~Brightwell, F.~Dowker, R.S. Garcia, J.~Henson, and R.D. Sorkin.
\newblock {''Observables'' in causal set cosmology}.
\newblock {\em Phys. Rev.}, D67:084031, 2003, [arXiv:gr-qc/0210061].

\bibitem{BDGHS01}
G.~Brightwell, H.F. Dowker, R.S. Garc\'{\i}a, J.~Henson, and R.D. Sorkin.
\newblock {General covariance and the 'Problem of Time' in a discrete
  cosmology}.
\newblock pages 1--17.
\newblock in K.G. Bowden, Ed., ''Correlations'', Proceedings of the ANPA 23
  conference, 2001, Cambridge, England (Alternative Natural Philosophy
  Association, London, 2002), [arXiv:gr-qc/0202097].

\bibitem{DG93}
M.~Droste and R.~G{\"o}bel.
\newblock {Universal domains and the amalgamation property}.
\newblock {\em Math. Struct. in Comp. Science}, 3:137--159, 1993.

\bibitem{DK03}
M.~Droste and D.~Kuske.
\newblock {On random relational structures}.
\newblock {\em Journal of Combinatorial Theory}, A 102:241--254, 2003.

\bibitem{ER63}
P.~Erd{\"o}s and A.~R\'{e}nyi.
\newblock {Asymmetric graphs}.
\newblock {\em Acta Math. Acad. Sci. Hungar.}, 14:295--315, 1963.

\bibitem{Fra54}
R.~Fraiss\'e.
\newblock {Sur l'extension aux relations de quelques propri\'et\'es des
  ordres}.
\newblock {\em Ann. Sci. \'Ecole Norm. Sup.}, 71:363--388, 1954.

\bibitem{G03}
Nicholas Georgiou.
\newblock {A random binary order: a new model of random partial orders}.
\newblock CDAM research report, London School of Economics, U.K., 2003.

\bibitem{H93}
W.~Hodges.
\newblock {\em {Model Theory}}.
\newblock Cambridge University Press, Cambridge, 1993.

\bibitem{MORS01}
X.~Martin, D.~O'Connor, D.~Rideout, and R.D. Sorkin.
\newblock {On the ''renormalisation'' transformations induced by cycles of
  expansion and contraction in causal set cosmology}.
\newblock {\em Phys. Rev.}, D63:084006, 2001, [arXiv:gr-qc/0009063].

\bibitem{Rad64}
R.~Rado.
\newblock {Universal graphs and universal functions}.
\newblock {\em Acta Arith.}, 9:331--340, 1964.

\bibitem{R01}
D.D. Reid.
\newblock {Discrete quantum gravity and causal sets}.
\newblock {\em Canadian Journal of Physics}, 79:1--16, 2001,
  [arXiv:gr-qc/9909075].

\bibitem{RS00}
D.~Rideout and R.~Sorkin.
\newblock {A classical sequential growth dynamics for causal sets}.
\newblock {\em Phys. Rev.}, D61:024002, 2000, [arXiv:gr-qc/9904062].

\bibitem{S02}
R.D. Sorkin.
\newblock {Causal Sets: Discrete Gravity}.
\newblock in the proceedings of the Valdivia Summer School, Valdivia, Chile,
  2002, edited by A. Gomberoff and D. Marolf (to appear),
  [arXiv:gr-qc/0309009].

\end{thebibliography}

\end{document}